\documentclass[twocolumn,prb,aps]{revtex4}

\usepackage{amsmath}
\usepackage{bm}
\usepackage{graphicx}
\usepackage{dcolumn}

\def\dfrac#1#2{{\displaystyle{\frac{#1}{#2}}}}

\begin{document}

\title{Intrinsic electric polarization in spin-orbit coupled semiconductor heterostructures}

\author{A.~V.~Rodina$^{1}$\cite{EMAIL} and A.~Yu.~Alekseev$^{2,3}$}

\affiliation{$^{1}$A. F. Ioffe Physico-Technical
Institute, 194021, St.-Petersburg, Russia\\
$^{2}$Department of Mathematics, University of Geneva, 1211 
Geneva, Switzerland\\
$^{3}$Institute of Theoretical Physics, Uppsala University,
S-75108, Uppsala, Sweden }
\date{\today}

\begin{abstract}

We present Maxwell equations with source terms for the electromagnetic field
interacting with a moving electron in a spin-orbit coupled  semiconductor heterostructure.  
We start
with the  eight--band ${\bm k}{\bm p}$ model and derive the electric and magnetic polarization
vectors using
the Gordon--like decomposition method. Next, we  present the ${\bm k}{\bm p}$
effective Lagrangian for the nonparabolic conduction band electrons interacting with
electromagnetic field  in semiconductor heterostructures with abrupt  
interfaces.
This Lagrangian gives rise to the Maxwell equations with source terms  
and boundary conditions at heterointerfaces as well as equations for  
the electron envelope wave function in the external electromagnetic  
field  together with appropriate boundary conditions.  As an example,  
we consider spin--orbit effects caused by the structure inversion  
asymmetry for the conduction electron states. We compute the intrinsic contribution 
 to the electric polarization of the steady state electron gas in asymmetric quantum well in  
equilibrium and in the spin Hall regime.
We argue that this contribution, as well as the intrinsic spin Hall current, 
are not cancelled by the elastic scattering processes.

\end{abstract}

\maketitle

\section{Introduction}

 Spintronic is a rapidly developing and important field of condensed  
matter physics. The research  is mainly concentrated on effects of  
electron spin transport, spin accumulation and spin manipulation in  
non-symmetric semiconductor heterostructures with strong  spin--orbit  
coupling. The early predictions\cite{DP1,DP2,hirsch} and recent  
experimental observations of the spin-Hall  
effect\cite{hallexp1,hallexp2} have inspired  a huge number of theory  
papers.\cite{universal,rashba,inoue,dimitrova,sohall,def,proper,korenev,consistency,gordon,dirak}
This research is primarily concentrated on the spin Hall current which  
is a flux of curriers with opposite spins in opposite directions  
perpendicular to the driving electric field. This current can be generated, for a example, due to the asymmetric scattering,\cite{DP1,DP2,hirsch} due to the diffusion of the nonequilibrium spin\cite{korenev,KKM} or  due to momentum dependent spin-orbit splitting in the band structure.\cite{universal,rashba,sohall} The latter effect is usually called   
the intrinsic effect (as it is computed with an equilibrium distribution function) and it is characterized by the universal spin Hall conductivity.\cite{universal,rashba,sohall}   

Several fundamental questions  concerning the spin Hall  
effect have inspired a wide discussion in the literature.
It concerns the definition of the electron spin current,\cite{def,proper,consistency,gordon,korenev} and the issue of spin Hall current cancellation in 
the steady state regime.\cite{rashba,inoue,dimitrova,korenev}
A good basis for treating these issues\cite{consistency,gordon,dirak} 
is the relativistic Dirac equation for an electron  
interacting with the electromagnetic
field. This approach reveals a close relation between the spin current and the electric  
polarization vector. It also gives a new contribution to the spin transfer torque 
coming from the interaction between the intrinsic electric polarization 
and the external electric field.\cite{consistency}

In the present paper, we describe an electron
interacting with the electromagnetic field and moving in a  
semiconductor heterostructures with strong spin-orbit coupling. We  
start with the eight-band  ${\bm k}{\bm p}$ Kane model  and  derive  
the expressions for the  electric and magnetic polarization  
vectors as well as the effective ${\bm k}{\bm p}$ Lagrangian  for the  
conduction band electrons in a semiconductor heterostructure with  
abrupt interfaces. Using the least action principle,  we  derive  
Maxwell equations with source terms and boundary conditions for the  
electromagnetic field at the interfaces as well as  equations for  
the electron envelope wave function in external fields together  with  
appropriate boundary conditions. As an example, we consider  
the steady state of the electron gas in an asymmetric quantum well. We compute the  
intrinsic spin-orbit contribution to the electric polarization in  
equilibrium, and in the spin Hall regime. We argue that the intrinsic electric polarization
in the spin Hall regime corresponds to an additional spin--orbit energy in the external electric field, and therefore it can not be cancelled by extrinsic contributions. Furthermore,
the intrinsic spin Hall current does not vanish in the steady state regime while the vanishing of the spin torque\cite{korenev} is maintained by additional contributions coming from  interaction
between the equilibrium electric polarization and the external electric field.\cite{consistency} We also predict an existence  of the intrinsic induced magnetic charge Hall current due to the inhomogeneous charge distribution in asymmetric quantum well. 

The paper is organized as follows: in Section \ref{dirac} we review  
some standard features of  the Dirac equation. In Sec. \ref{kane} we   
review the properties of the eight-band  ${\bm k}{\bm p}$ Kane model 
 which includes an external electromagnetic field and    
derive the expressions for the electric and magnetic polarization vectors.  In Sec.  
\ref{lap} we derive the effective ${\bm k}{\bm p}$ Lagrangian for the   
conduction band electron  interacting with  external electromagnetic field 
and apply the least action principle to  
semiconductor heterostructures. In Sec. \ref{sample} we calculate the  
intrinsic spin--orbit contribution to the electric polarization  of the steady state electron gas   in an asymmetric quantum  
well in  equilibrium and  spin Hall regimes as well as  the intrinsic induced magnetic charge Hall current. In Sec. \ref{discuss} we discuss the results and their relevance for the
fundamental issues  of the proper definition of the spin Hall current, and of its non 
cancellation in the steady state regime. 

\section{Electromagnetic polarization induced by Dirac electron}
 \label{dirac}

We start with the problem  of a relativistic electron interacting with the electromagnetic field in the vacuum.  
Maxwell equations for the electric field ${\bm E}$ and magnetic induction ${\bm B}$  can be written (in SGS units) as:
\begin{eqnarray}
{\bm \nabla}\cdot {\bm E} = 4\pi\rho  \, ,  \quad {\bm \nabla}\times {\bm B}= \frac{1}{c}\frac{\partial {\bm E}}{\partial t}  + \frac{4\pi}{c} {\bm J}\,, \label{max1} \\
{\bm \nabla}\cdot {\bm B} = 0 \, , \quad 
{\bm \nabla}\times {\bm E}= - \frac{1}{c} \frac{\partial {\bm B}}{\partial t}  \, ,
 \label{max2}
\end{eqnarray}
where $c$ is the velocity of light. The charge density $ \rho$ and the current density ${\bm J}$ correspond to a single moving electron and satisfy the continuity equation:
\begin{equation}
\frac{\partial \rho }{\partial t} + {\bm \nabla}\cdot {\bm J} = 0 \, .
\label{cont} 
\end{equation}
For a Dirac electron, $\rho=e\Psi^*\Psi $ and ${\bm J}=ce(\Psi^*{\bm \alpha }\Psi)$, where $e=-|e|$ is  the free electron charge and  $\Psi$ is a four component (bispinor) wave function satisfying the Dirac equation
\begin{eqnarray}
\left( i\hbar \frac{\partial }{\partial t} - eV \right)\Psi  =(c{\bm \alpha}\cdot {\bm \pi } 
+mc^2\beta )\Psi \, , \\
{\bm \alpha }=
 \left(
 \begin{array}{cc} 0 & {\bm \sigma } \\ {\bm \sigma } & 0
\end{array} \right), \quad  \beta=  \left(\begin{array}{cc} \hat 1 & 0 \\ 0 & - \hat 1 \end{array} \right) , \quad 
\hat 1=  \left(\begin{array}{cc} 1 & 0 \\ 0 & 1 \end{array} \right) \, .
\label{dir}
\end{eqnarray}
Here $m$ is the  free electron mass, $ \hat {\bm \sigma}= \{ \hat \sigma_x ,
\hat \sigma_y ,  \hat \sigma_z \}$  are the Pauli matrices,
 ${\bm \pi }= -i\hbar {\bm \nabla} - (e/c) {\bm A}$ is the momentum operator,
 and $V$ and ${\bm A}$ are scalar and vector potentials of  the electromagnetic fields ${\bm E}$ and ${\bm B}$, respectively:  
\begin{eqnarray}
{\bm E} = -(\partial {\bm A}/\partial t) - 
{\bm \nabla} V\, , \qquad 
 {\bm B} = {\bm \nabla} \times {\bm A}.
 \end{eqnarray}

The Gordon decomposition is a way to split $\rho$ and ${\bm J}$ 
  into  convective and internal parts,\cite{gordon}
$\rho =\rho_c + \rho_i$, ${\bm J}={\bm J}_c + {\bm J}_i$.  The convective parts are
given by 
\begin{eqnarray}
\rho_c= \frac{i\hbar }{2mc^2} \left(\bar \Psi \frac{\partial \Psi }{\partial t} -  
\frac{\partial \bar \Psi }{\partial t} \Psi \right) - \frac{eV}{mc^2}
{\bar  \Psi} \beta \Psi   \, , \\
{\bm J}_c= \frac{e}{2m} \left[ \bar \Psi  ({\bm \pi } \Psi ) +
 ({\bm \pi}^* \bar \Psi )\Psi \right] \, ,
\end{eqnarray} 
where $\bar \Psi = \Psi^*\beta$.  
 The internal parts have the form 
\begin{equation}
\rho_i = - {\bm \nabla} \cdot {\bm \Pi} \, , \quad 
{\bm J}_i=c {\bm \nabla} \times {\bm M} + \frac{\partial \bm \Pi}{\partial t} \, ,
\end{equation}
where  electric and magnetic polarizations ${\bm \Pi}$ and ${\bm M}$ are given by
\begin{eqnarray}
&&{\bm \Pi} =\frac{ e \hbar }{2mc} \bar \Psi (-i{\bm \alpha  }) \Psi  \, ,\quad
\nonumber \\&&{\bm M} =\frac{e \hbar }{2mc} \bar \Psi {\bm \Sigma } \Psi  \, ,\quad 
{\bm \Sigma }=
 \left(
 \begin{array}{cc}  {\bm \sigma } & 0 \\ 0 & {\bm \sigma } 
\end{array} \right) \, . 
\label{pim}
\end{eqnarray}
The convective and internal densities are separately conserved:
$
{\partial \rho_{c,i} }/{\partial t} + {\bm \nabla}\cdot {\bm J}_{c,i} = 0 \, .
$
 The internal density $\rho_i  $ and current ${\bm J}_i$ are the densities of the induced electric charge and the induced electric current of a moving electron. One can then rewrite the  first pair of Maxwell equations (\ref{max1}) as  
\begin{eqnarray}
{\bm \nabla}\cdot {\bm D} = 4\pi \rho_c  \,  , \quad 
{\bm \nabla}\times {\bm H}= \frac{1}{c} \frac{\partial {\bm D}}{\partial t} 
 + \frac{4\pi}{c}{\bm J}_c\, ,
\label{max22}
\end{eqnarray}
where ${\bm D}= {\bm E} + 4\pi {\bm \Pi} $ is the electric displacement vector and ${\bm H} = {\bm B} - 4\pi {\bm M}$ is the magnetic field strength in the vacuum. In the general case of dielectric and magnetic media, vectors ${\bm D}$ and ${\bm H}$ are related to vectors ${\bm E}$ and ${\bm B}$ via
\begin{eqnarray}
{\bm D} =  \varepsilon {\bm E} + 4\pi {\bm \Pi}  \, ,\qquad 
{\bm H}={\bm B}/\mu - 4\pi {\bm M} \, .
\label{DH}
\end{eqnarray}
Here $\varepsilon$ and $\mu$ are electric permittivity  and magnetic conductivity   of the media, respectively,  and vectors ${\bm \Pi}$ and ${\bm M}$ describe polarizations induced by a moving Dirac electron.

Equation ${\partial \bm \Pi}/{\partial t}={\bm J}- {\bm J}_c-c {\bm \nabla} \times {\bm M}$ is a direct consequence of the Dirac equation (\ref{dir}) and its complex conjugate. A similar calculation leads to the following equation for the magnetic polarization vector:
\begin{eqnarray}
  \frac{\partial \bm M}{\partial t} =&{\bm J}_m + c {\bm \nabla} \times {\bm \Pi} \, .
\label{mpi}
\end{eqnarray}
Here 
\begin{eqnarray}
 {\bm J}_m & =&  - \, \frac{ie}{2m} \left( ({\bm \pi}^*\bar \Psi {\bm \alpha}){\bm \Sigma}\Psi + \bar \Psi {\bm \Sigma} ({\bm \alpha}{\bm \pi} \Psi) \right) - c {\bm \nabla} \times {\bm \Pi} = \nonumber \\
 &=& \frac{e}{2m} \left[ \Psi^* \rho_2  ({\bm \pi } \Psi ) +
 ({\bm \pi}^*  \Psi^* \rho_2 )\Psi \right] , \quad \\
&&\rho_2 \equiv   \left(\begin{array}{cc} 0& -i \hat 1 \\i \hat 1 & 0 \end{array} \right) \nonumber
\end{eqnarray}
is the induced magnetic charge current. It satisfies the continuity equation 
\begin{eqnarray}
\frac{\partial \rho_{m} }{\partial t} + {\bm \nabla}\cdot {\bm J}_{m} = 0 \, ,
\end{eqnarray} 
where 
$
\rho_m = - {\bm \nabla} \cdot {\bm M} \, 
$
is the induced magnetic charge density.  

Now Maxwell equations for the displacement vectors ${\bm D}$ and ${\bm H}$ can be   written in the symmetric form: 
\begin{eqnarray}
{\bm \nabla}\cdot {\bm D} = 4\pi \rho_c  \, , \label{max1d} \\
{\bm \nabla}\cdot {\bm H} = -4\pi \rho_m  \, , \label{max1h} \\
{\bm \nabla}\times {\bm D}= -\frac{1}{c}\frac{\partial {\bm H}}{\partial t} - \frac{4\pi}{c}{\bm J}_m \, , \label{max2d} \\
{\bm \nabla}\times {\bm H}= \frac{1}{c}\frac{\partial {\bm D}}{\partial t} 
 + \frac{4\pi}{c}{\bm J}_c\, \label{max2h} .
\end{eqnarray}
These equations are similar to those presented in Ref. \onlinecite{elfield}. 
Note that the induced magnetic charge density $\rho_m$ and the corresponding
current density ${\bm J}_m$ also appear in the classical electrodynamics
of moving media.\cite{move}

For a Dirac electron, expressions (\ref{pim}) for ${\bm \Pi}$ and  ${\bm M}$ are exact.  Approximate expressions for ${\bm \Pi}$ and ${\bm M}$ in weakly relativistic limit  
were recently obtained in Ref. \onlinecite{consistency}.  However, the Dirac equation 
or its weakly relativistic limit can not be  directly applied to  the case of semiconductor heterostructures with strong spin--orbit interaction.\cite{rwink} 
In the next section, we start with the eight--band ${\bm k}\cdot {\bm p}$ Kane model,
and we derive expressions for  electric and magnetic polarizations induced by the spin--orbit coupling of conduction band electrons.

\begin{widetext}
\section{Kane electron in external electromagnetic field }
\label{kane}

The energy band structure of cubic semiconductors near the
center of the first Brillouin zone can be described within the
eight--band $\bm{k \cdot p}$  model.\cite{bir,ivchenko} In
 homogeneous bulk  semiconductor, the  full wave function  can be
 expanded as\cite{bir}
\begin{eqnarray}
{\Psi ({\bm r})} =\sum_{\mu = \pm 1/2}
\Psi_{c}^\mu({\bm r}) |S\rangle
u_{\mu} \, +
\, \sum_{ \mu = \pm 1/2  } \sum_{\alpha=x,z,z}  \Psi_{v\,
\alpha}^\mu({\bm r}) |R_\alpha \rangle u_{\mu} \,
,\label{psi}
\end{eqnarray}
where  $u_{1/2}$ and $u_{-1/2}$ are the eigenfunctions of the spin
operator $\hat {\bm S} = (\hbar/2) \, \hat
{\bm \sigma}$. 
$|S\rangle $ is the Bloch function of the conduction band edge at the
$\Gamma$--point of the Brillouin zone which represents an eigenfunction of
internal momentum $I=0$. $|R_x\rangle\equiv |X\rangle$,
$|R_y\rangle\equiv |Y\rangle$, $|R_z\rangle\equiv |Z\rangle$ are Bloch
functions of the valence band edge at the $\Gamma$--point of
the Brillouin zone. Combinations of these functions
$(|R_x\rangle \pm i |R_y\rangle)/\sqrt{2}$ and $|R_z\rangle$  are
eigenfunctions of the internal momentum $I=1$ with projections
on the $z$ axis equal to $\pm 1$ and $0$, respectively (see Ref.
\onlinecite{bir}).  Smooth functions $\Psi_{c}^{\pm
1/2}({\bm r})$ are components of the conduction band
spinor envelope function $\Psi_c=\left( \begin{array}{c} \Psi_c^{1/2}
\\ \Psi_c^{-1/2} \end{array} \right)$, and $\Psi_{v\, x}^{\pm
1/2}({\bm r})$, $\Psi_{v\, y}^{\pm 1/2}({\bm r})$,
 $\Psi_{v\,z}^{\pm 1/2}({\bm r})$ are  $x,y,z$
components of the valence  band spinor envelope vector
$${\bm \Psi}_v=\left( \begin{array}{c} {\bm \Psi}_v^{1/2} \\
 {\bm \Psi}_v^{-1/2} \end{array} \right)=\left\{ \left(
\begin{array}{c} \Psi_{v\,x}^{1/2} \\ \Psi_{v\,x}^{-1/2} \end{array}
\right),
\left( \begin{array}{c} \Psi_{v\,y}^{1/2}\\
\Psi_{v\,y}^{-1/2} \end{array} \right),
\left( \begin{array}{c}
\Psi_{v\,z}^{1/2} \\ \Psi_{v\,z}^{-1/2} \end{array} \right)\right\} \, .$$

\subsection{Basic equations}

In the bulk,  the eight--component
envelope function ${\Psi} ({\bm r}) \equiv \{
{\Psi}_c({\bm r}) ,{\bm \Psi}_v({\bm r}) \}$ is a solution of the Schr\"{o}dinger equation\cite{polk,so,book}
\begin{eqnarray}
i\hbar \frac{\partial }{\partial t}\left(\begin{array}{c} \Psi_c \\
   {\bm \Psi}_v \end{array} \right)\, &=& \hat H_{\rm Kane} \left(\begin{array}{c} \Psi_c \\
   {\bm \Psi}_v \end{array} \right)\,  , \label{psifull} \\ 
\hat H_{\rm Kane} \left(\begin{array}{c} \Psi_c  \\
   {\bm \Psi}_v \end{array} \right)\,  &=& 
  \left(  \begin{array}{cc}  \dfrac{\alpha \hbar
^{2}}{2m}  \hat k^{2}  \Psi_c&
     i  P\hbar  (\hat {\bm k} {\bm \Psi}_v )\\
- i  P \hbar \hat {\bm k} \Psi_c &
(-E_g - \dfrac{\Delta}{3}){\bm \Psi}_v + \dfrac{i\Delta}{3}{\bm \sigma} \times {\bm \Psi}_v 
\end{array} \right) \,  . \label{hfull}
\end{eqnarray}
Here the energy of  electron states is measured with respect to the bottom of the conduction
band $E_c=0$,  $E_g=E_c-E_v$ is the band gap energy, $\Delta$ is the spin--orbit splitting of the valence band, $ \hat {\bm  k}= - i {\bm \nabla} $ is the wave vector, and $P= -i\hbar \langle S|\hat{p}_z|Z\rangle/m$
is  the Kane matrix element describing the coupling of the conduction
and valence bands.  The parameter $\alpha$ describes the contribution to  the
 electron effective mass $m_c$ which is not related to the interaction with the valence band, while the $k^2$ terms for the valence band are neglected. This is the so called the eight-band Kane model with dispersion for electrons only.\cite{polk,so,book} This model allows to describe electron states with energies in the conduction band. It takes into account  spin--orbit effects induced by the interaction with the valence band in the presence of the structure inversion asymmetry.\cite{so}  Bulk inversion asymmetry terms are  not included in the consideration.

We introduce an expression 
\begin{eqnarray}
{\bm\Psi}_l = -i\hbar \frac{\partial {\bm \Psi}_v}{\partial t} - \hbar P {\bm \nabla} \Psi_c \, , 
\label{psil}
\end{eqnarray}
and rewrite  
the second vector equation of (\ref{psifull},\ref{hfull}) as 
\begin{eqnarray}
{\bm\Psi}_l  = \left( E_g + \frac{\Delta}{3} \right){\bm \Psi}_v - \frac{i\Delta}{3} {\bm \sigma} \times {\bm \Psi}_v \, \label{psiv1} \, .
\end{eqnarray}
Taking the cross product with  ${\bm \sigma}$, we obtain an equation  for ${\bm \sigma} \times {\bm\Psi}_l$:  
\begin{eqnarray}
{\bm \sigma} \times {\bm\Psi}_l = \left( E_g + \frac{2\Delta}{3} \right){\bm \sigma} \times{\bm \Psi}_v + \frac{2i\Delta}{3}{\bm \Psi}_v \, .
\label{psiv2}
\end{eqnarray}
Combining (\ref{psiv1}) and (\ref{psiv2}), we express 
the 
valence band  spinor vector ${\bm \Psi}_v$ and the vector product ${\bm \sigma} \times {\bm \Psi}_v$ as 
\begin{eqnarray}
 {\bm \Psi}_v =  C_1 {\bm \Psi}_l + i C_2 [\hat \sigma \times {\bm\Psi}_l] \, , 
\label{psiv} \\
{\bm \sigma} \times {\bm \Psi}_v =  -2i C_2 {\bm \Psi}_l + (C_1- C_2) [\hat \sigma \times {\bm\Psi}_l] \, .
\label{psis}
\end{eqnarray}
Here the coefficients $C_1 $ and $C_2$ are given by 
\begin{equation}
C_1 \equiv  \frac{3E_g+2\Delta}{3E_g(E_g + \Delta)} \, , \quad 
C_2 \equiv  \frac{\Delta}{3E_g(E_g + \Delta)} \, .
\label{c12}
\end{equation}
They  are related to the electron effective mass $m_c$ and the electron effective $g$-factor $g_c$ at the bottom of the conduction band via
\begin{eqnarray}
\frac{m}{m_c} = \alpha + E_p C_1 \, , \quad g_c = g_e - 2E_p C_2 \, .
\end{eqnarray}
Here $E_p=2mP^2$ is the Kane energy parameter, $g_e=g_0+g^{*}$, $g_0 \approx 2$ is the free electron $g$ factor  and $g^{*}$ describes the remote band contribution. 

To include the interaction with the electromagnetic field we replace $\hbar {\bm k}= -i \hbar {\bm \nabla}$ with ${\bm \pi} =  -i \hbar {\bm \nabla} - e/c {\bm A}$ and $i\hbar\, \partial/\partial t$ with  $i\hbar\, \partial/\partial t -eV$ in Eqs. (\ref{psifull},\ref{hfull},\ref{psil}), and we add the respective  Zeeman Kane Hamiltonian $\hat H_{\rm Zeeman}$:\cite{so}
 \begin{eqnarray}
\hat H_{\rm Zeeman} \left(\begin{array}{c} \Psi_c  \\
   {\bm \Psi}_v \end{array} \right)\,  = 
  \left(  \begin{array}{cc}  \frac{1}{2} g_e \mu_B ({\bm B}{\bm \sigma})  \Psi_c&
    0 \\ 0 &
 \frac{1}{2} g_0 \mu_B ({\bm B}{\bm \sigma})  {\bm \Psi}_v + i \mu_B {\bm B} \times {\bm \Psi}_v 
\end{array} \right) \,  . \label{hZeeman}
\end{eqnarray}
Here $\mu_B= |e|\hbar/2mc$ is the Bohr magneton. In order to simplify our consideration, we assume that the only spin--orbit contributions come from the interaction between conduction band and  valence band states. Hence, the Hamiltonian does not include any Rashba terms related to the remote band contributions. 

In the presence of an additional Zeeman Hamiltonian (\ref{hZeeman}) the decompositions  (\ref{psiv},\ref{psis}) are not exact. To take into account  first order corrections coming from  (\ref{hZeeman})  one has to replace ${\bm \Psi}_l$ with ${\bm \Psi}_l+ {\bm \Psi}_B$ in Eq. (\ref{psiv},\ref{psis}). Here the correction term
\begin{eqnarray}
{\bm \Psi}_B \approx \mu_B({\bm \sigma}{\bm B})(C_1 {\bm \Psi}_l + i C_2 [\hat \sigma \times {\bm\Psi}_l]) + i\mu_B {\bm B} \times (C_1 {\bm \Psi}_l + i C_2 [\hat \sigma \times {\bm\Psi}_l] )   \, 
\label{psib}
\end{eqnarray}
is by a factor of $\mu_B|B|/E_g$ smaller than  ${\bm \Psi}_l$, and it can usually be neglected.

\subsection{Electron steady state in the stationary electromagnetic field}

We would like to compute the electron steady state with energy ${\cal E}$ in the stationary electromagnetic field. To this end, we replace  $i\hbar \partial \Psi/\partial t$ with  $ ({\cal E}-eV) \Psi$, where $V$ is the scalar potential: ${\bm E} = - {\bm \nabla} V$. Then, the valence band contribution takes the form
\begin{eqnarray}
{\bm \Psi}_v(\epsilon ) = {\bm \Psi}_\epsilon  + {\bm \Psi}_B \, ,  \label{psive1} \\ 
{\bm \Psi}_\epsilon =  -i P C_1(\epsilon ) {\bm \pi} \Psi_c +  P C_2(\epsilon ) [\hat \sigma \times {\bm \pi} \Psi_c] \, , \label{psive2} \\
{\bm \Psi}_B \approx \mu_B({\bm \sigma}{\bm B}){\bm \Psi}_\epsilon  + i\mu_B {\bm B} \times {\bm \Psi}_\epsilon   \, .
\label{psive3}
\end{eqnarray}
Here $\epsilon={\cal E}-eV$, and  the coefficients $C_1({\cal E}-eV)$ and  $C_2({\cal E}-eV)$ coincide with coefficients $C_1$ and $C_2$ given by Eq. (\ref{c12}) after replacing $E_g$ by $E_g +{\cal E} -eV$. The resulting non-parabolic equation for the conduction band spinor function reads: 
\begin{eqnarray}
\left( {\bm \pi}\frac{1}{2m_c(\epsilon)}{\bm \pi} + i\, {\bm \pi} \frac{(g_c(\epsilon)-g_e)}{4m} [{\bm \sigma} \times {\bm \pi}] +
\frac{\mu_B}{2}g_e({\bm \sigma}{\bm B}) \right) \Psi_c + iP{\bm \pi} \Psi_B = \epsilon \Psi_c \, , 
\label{psic1}
\end{eqnarray}
where again $\epsilon={\cal E}-eV$. The energy dependent electron effective mass $m_c(\epsilon )$ and $g$--factor $g_c(\epsilon )$ are given by:
\begin{eqnarray}
\frac{m}{m_c(\epsilon) } = \alpha + E_p C_1(\epsilon)  \, , \quad g_c(\epsilon)  = g_e - 2E_p C_2(\epsilon)  \, .
\end{eqnarray}
The contribution of $\Psi_B$ in Eq. (\ref{psive3}) [the last term on the left hand side of Eq. (\ref{psic1})] gives corrections to the first two terms which are proportional to a small factor $|\mu_B B|/E_g$, and usually it can be neglected. 

For  small energies  $|\epsilon| \ll E_g$, the nonparabolic electron effective mass and electron effective $g$-factor can be expanded near the bottom of the conduction band
 \begin{eqnarray}
{m\over m_c(\epsilon )}={m\over m_c}- {\alpha_p}\epsilon \, ,\qquad 
g_c(E)=g_c+{\alpha_{so}}\epsilon  \, , 
\label{mpg}
\end{eqnarray}
where $\alpha_p$ is the mass nonparabolicity parameter and 
$\alpha_{so}$ is the g-factor nonparabolicity parameter closely related to the spin-orbit coupling constant:\cite{rwink,silva}
\begin{eqnarray}
a_p= \frac{E_p}{3}\left[ \frac{2}{E_g^2} + \frac{1}{(E_g+\Delta)^2} \right] = E_p ( C_1^2 + 2C_2^2) \, , \\
\alpha_{so}= {2E_p\over 3}\left[{1\over E_g^2} -{1\over
(E_g+\Delta)^2}\right] = 2E_pC_2(2C_1-C_2) \, .
\end{eqnarray}
Substituting the expansion of (\ref{mpg}) into (\ref{psic1}) and neglecting the contribution of $\Psi_B$ of Eq. (\ref{psive3}), we arrive at 
\begin{eqnarray}
\left( {\bm \pi}\frac{1}{2m_c(\epsilon)}{\bm \pi} + \frac{e \hbar}{4m} \alpha_{so} {\bm E} [{\bm \sigma} \times {\bm \pi}] +
\frac{\mu_B}{2}g_c(\epsilon) ({\bm \sigma}{\bm B}) \right) \Psi_c = \epsilon \Psi_c \, . 
\label{psic2}
\end{eqnarray}
Note that  $\Psi_c$ in Eqs. (\ref{psic1}) and (\ref{psic2}) is the original conduction band spinor, and hence the  normalization condition  reads
$
\int (|\Psi_c|^2+(|\Psi_v|^2){\bm d}^3{\bm r} =1. 
$
Expressing ${\bm \Psi}_v$ via $\Psi_c$ with the help of Eqs. (\ref{psive1}, \ref{psive2}, \ref{psive3}) and keeping only the first order terms in $|\epsilon|/E_g$ and $|\mu_BB|/E_g$, we obtain the approximate  normalization condition for $\Psi_c$
\begin{eqnarray}
\int \left( |\Psi_c|^2 -  \frac{\hbar^{2}a_p}{4m} 
\left( \Psi_c^* {\bm \nabla}^2 \Psi_c + {\bm \nabla}^2 \Psi_c^* \Psi_c \right) + \frac{ \alpha_{so}}{2} \mu_B \Psi_c^*({\bm \sigma B}) \Psi_c \right) {\bm d}^3{\bm r} = 1 \, .
\end{eqnarray}
In  absence of  external fields, this condition can be presented  as
\begin{eqnarray}
\int  |\Psi_c|^2 {\bm d}^3{\bm r} = \frac{m_c}{m_c({\cal E})} \, .
\end{eqnarray}

\subsection{Electric and magnetic polarizations in the Kane model}

The continuity equation (\ref{cont}) for the charge density $\rho$ and the electric current density ${\bm J}$ in the  Kane model can be obtained directly from the Schr\"{o}dinger equation (\ref{psifull}). 
A straightforward calculation leads to the following expressions:
\begin{eqnarray}
&&\rho =e(|\Psi_c|^2 + |{\bm \Psi}_v|^2) \, , \label{rho} \\
&&{\bm J}=\frac{ e \alpha}{2m}( ({\bm \pi} \Psi_c)^*\Psi_c  + \Psi_c^* {\bm \pi} \Psi_c)\, + 
 {ieP}\left( \Psi_c^*{\bm \Psi}_v - {\bm \Psi}_v^*\Psi_c \right)
+ c{\bm \nabla} \times {\bm M}_0 \, , \label{j} \\
&&{\bm M}_0 = -\frac{\mu_Bg_e}{2} \Psi_c^* {\bm \sigma} \Psi_c  -\frac{\mu_Bg_0}{2}\sum_{\gamma=x,y,z}\Psi_{v  \gamma}^* {\bm  \sigma} \Psi_{v \gamma}\,  - \,i \,\mu_B [ {\bm \Psi}_{v}^* \times  {\bm \Psi}_{v}  ] \, . \label{m}
\end{eqnarray} 

Using the decomposition given by Eqs. (\ref{psiv}-\ref{psis}) and neglecting the contribution of ${\bm \Psi}_B$ in Eq. (\ref{psib}) one can separate   convective and internal parts of $\rho$ and ${\bm J}$ as 
$\rho=\rho_c + \rho_i$ and ${\bm J}={\bm J}_c + {\bm J}_i$, where internal parts $\rho_i$ and ${\bm J}_i$ are related to electric and magnetic polarization vectors ${\bm P}$ and ${\bm M}$ via Eq. (9). For the Kane model, we obtain the following  approximate expressions for $\rho_c$, ${\bm J}_c$, ${\bm \Pi}$ and ${\bm M}$: 
\begin{eqnarray}
&&\rho_c=e|\Psi_c|^2 +\frac{e\hbar P}{2}
\left( \Psi_c^* ({\bm \nabla}\bar {\bm \Psi}) - ({\bm \nabla}\bar {\bm \Psi}^*) \Psi_c \right) \, +  \\
&&+ \frac{ie\hbar}{2} \left( \frac{\partial {\bm \Psi}_v^*}{\partial t} \bar {\bm \Psi} - \bar {\bm \Psi}^* 
\frac{\partial {\bm \Psi}_v}{\partial t} \right) + \frac{e^2V}{2} \left( {\bm \Psi}_v^* \bar {\bm \Psi} +
\bar {\bm \Psi}^* {\bm \Psi}_v \right) \, , \nonumber \\
&&{\bm \Pi}=\frac{e\hbar P}{2} \left( \Psi_c^* \bar {\bm \Psi} +  \bar {\bm \Psi}^*\Psi_c \right) \, , \qquad \bar {\bm \Psi} = C_1{\bm  \Psi}_v + i C_2 {\bm \sigma} \times {\bm \Psi}_v  \, ,\\
&&{\bm J}_c = \frac{e}{2m_c(-eV)}\left(\Psi_c^*{\bm \pi}\Psi_c + ({\bm \pi} \Psi_c)^*\Psi_c \right) + 
    \frac{\alpha_{so}e^2\hbar}{2m} \Psi_c^*({\bm \nabla} V \times {\bm \sigma} )\Psi_c \, + \\
&&+ \frac{eP\hbar}{2}\left( \Psi_c^* \frac{\partial \bar {\bm \Psi}}{\partial t} +  
\frac{\partial \bar {\bm \Psi}^*}{\partial t} \Psi_c - \frac{\partial  { \Psi}_c^*}{\partial t}\bar {\bm \Psi} - \bar {\bm \Psi}^*  \frac{\partial  { \Psi}_c}{\partial t} \right) \, , \nonumber \\
&&{\bm M} = -\frac{1}{2}\mu_B g_c(-eV ) \Psi_c^* {\bm \sigma} \Psi_c -\frac{\mu_Bg_0}{2}\sum_{\gamma=x,y,z}\Psi_{v  \gamma}^* {\bm  \sigma} \Psi_{v \gamma}\,  - \,i \,\mu_B [ {\bm \Psi}_{v}^* \times  {\bm \Psi}_{v}  ]  \, .
 \end{eqnarray} 
It is important to note that the convective charge density $\rho_c$ and current ${\bm J}_c$ enter Maxwell equations (\ref{max2}) for vectors ${\bm D}$ and ${\bm H}$ which are related to ${\bm E}$ and ${\bm B}$ by expressions (\ref{DH}). 
Vectors ${\bm \Pi}$ and ${\bm M}$ in Eq. (\ref{DH}) describe contributions of a moving Kane electron into the total electric and magnetic polarizations which are not taken into account by the material permittivity tensors $\varepsilon$ and $\mu$. 

Finally, for an electron steady state with  energy $|\epsilon|$ in a stationary magnetic field one can apply the decomposition given by  Eqs. (\ref{psive1}, \ref{psive2}, \ref{psive3})  to Eqs. (\ref{rho}, \ref{j}, \ref{m}) . Neglecting the contribution of $\Psi_B$ in Eq. (\ref{psive3}) and keeping only the first order terms in $|\epsilon|/E_g$ and $|\mu_BB|/E_g$,  we arrive at the final expressions  for the source terms:
\begin{eqnarray}
&&\rho_c = e|\Psi_c|^2 -  \frac{e\hbar^{2}a_p}{4m} 
\left( \Psi_c^* {\bm \nabla}^2 \Psi_c + {\bm \nabla}^2 \Psi_c^* \Psi_c \right) + \frac{ e\alpha_{so}}{2} \mu_B \Psi_c^*({\bm \sigma B}) \Psi_c \, , \label{s1} \\
&&{\bm \Pi} = - \frac{ e \hbar^{2} a_p}{4m} {\bm \nabla} |\Psi_c|^2  - 
\frac{e \alpha_{so}\hbar}{8m} \left( \Psi^* {\bm \sigma} \times {\bm \pi} \Psi_c - 
({\bm \pi} \Psi)^* \times {\bm \sigma} \Psi_c  \right) \, ,  \label{s2}\\
&&{\bm J}_c = \frac{e}{2m_c(\epsilon )}\left(\Psi_c^*{\bm \pi}\Psi_c + ({\bm \pi} \Psi_c)^*\Psi_c \right) + 
  \frac{\alpha_{so}e^2\hbar}{2m} \Psi_c^*({\bm \nabla} V \times {\bm \sigma} )\Psi_c \, ,  \label{s3}\\
&&{\bm M} = -\frac{1}{2}\, \mu_B g_c(\epsilon ) \Psi_c^* {\bm \sigma} \Psi_c \, .  \label{s4}
\end{eqnarray}
 
\end{widetext}

\section{Least-action principle for  semiconductor heterostructure in external electromagnetic field }
\label{lap}

In the previous section, we found the effective charge density, the electric current density and electric and magnetic polarization vectors associated to conduction band electrons.  The alternative approach is based on the Lagrangian formalism and on the ${\bm k}{\bm p}$ analog of the least action principle derived in 
Ref. \onlinecite{lap}. This approach is particularly efficient in applications to abrupt heterostructures. It has two main advantages: (i) the variation of the action provides equations of motion together with boundary conditions at heterointerfaces and (ii) incorporating of external electromagnetic field is straightforward.

The time-independent effective mass Lagrangian density for $\Gamma_6$ electrons with nonparabolicity is given by:\cite{lap}
\begin{eqnarray} 
{\cal L}({\cal E}) &= & {\cal E}|\Psi_c|^2 -
\frac{\hbar^2}{2m_c({\cal E})}|\nabla \Psi_c ({\bm r})|^2 + {\cal L}_{\rm SIA}({\cal E}) \, ,
\label{lema}  \\
{\cal L}_{\rm SIA}({\cal E})& = & - \frac{i\hbar^2}{4m}
(g_e-g({\cal E})) {\bm \nabla} \Psi_c^* [{\bm \sigma} \times {\bm \nabla} \Psi_c]
\, .
 \label{lesia} 
\end{eqnarray}
In the presence of external stationary electro-magnetic field,
the Lagrangian density is
\begin{equation}
{\cal L}_{\rm el-EM} =  {\cal L}_{\rm EM} + {\cal L}(\epsilon) + {\cal L}_{\rm Zeeman} \, . 
\label{L1}
\end{equation}
Here the Lagrangian density ${\cal L}_{\rm EM}$  of the stationary electromagnetic field takes into account material permittivity tensors $\varepsilon$ and $\mu$ characterizing the material properties in absence of moving electrons:
\begin{eqnarray}
{\cal L}_{\rm EM}= \frac{1}{8\pi} \left( {\bm E}_\alpha \varepsilon_{\alpha \beta} {\bm E}_\beta - {\bm B}_{\alpha} \frac{1}{\mu}_{\alpha \beta} {\bm B}_{\beta} \right) \, .
\label{L2}
\end{eqnarray}
The Lagrangian density ${\cal L}(\epsilon )$ can be obtained from Eqs. (\ref{lema},\ref{lesia}) by replacing ${\cal E}$ with $\epsilon={\cal E}-eV$ and $-i \hbar {\bm \nabla}$ with 
${\bm \pi} =  -i \hbar {\bm \nabla} - e/c {\bm A}$.  ${\cal L}_{\rm Zeeman}$ corresponds to the conduction band Zeeman energy  
\begin{eqnarray}
{\cal L}_{\rm Zeeman} = -\frac{\mu_B}{2} g_e \Psi_c^* ({\bm \sigma}{\bm B})\Psi_c  \, .
\label{L3}
\end{eqnarray}

\begin{figure}
\begin{center}
\includegraphics[width=7.5cm,height=7.5cm]{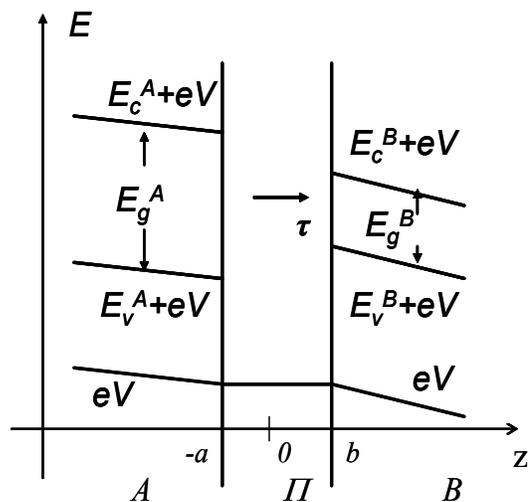}
\end{center}
\caption{\label{Fig1}  Sketch of  a planar heterointerface between
semiconductor layers $A$. $E_c^{A,B}+eV$ and $E_v^{A,B}+eV$ are the conduction band bottom  and the valence band top energies, respectively,  in the regions $A$ and $B$. These energies are not defined in the boundary region $\Pi$, while the scalar electromagnetic potential $V$ is continuous everywhere.}
\end{figure}

We consider  a planar semiconductor heterostructure consisting of two bulk--like regions $A$ and $B$ connected by a thin boundary region $\Pi$ around the abrupt heterointerface (see Fig. {\ref{Fig1}). The envelope function components $\Psi_c$ are defined only in the bulk--like regions of the heterostructure,\cite{bc,lap} and they obey  the boundary conditions at $z=-a$ and $z=b$. The material parameters $m_c$, $g_c$, $\alpha$, $g_e$, $E_p$ and $E_g$ may abruptly change from the region $A$ to the region $B$, and they are not defined in the boundary region $\Pi$. The scalar potential $V$ and the vector potential ${\bm A}$ are  continuous throughout the heterostructure and do not change inside the thin boundary region $\Pi$. The stationary fields ${\bm E} = -{\bm \nabla} V$ and ${\bm B}= {\bm \nabla} \times {\bm A}$ do not have any $\delta$-function components, and they are subject to the boundary conditions at  $z=-a$ and $z=b$. For the sake of simplicity, we assume that the material permittivity tensors $\varepsilon$ and $\mu$ are the same in $A$ and $B$. Then, the Lagrangian densities in the bulk--like regions $A$ and $B$ are given by Eqs. (\ref{lema}-\ref{L3}) with the material parameters of the materials $A$ and $B$, respectively. Note, that the energy $E$ of the electron steady state should be replaced with $E-E_c^{A,B}$ while $E+E_g$ should be replaced  with $E-E_v^{A,B}$. Here $E_c^{A,B}$ and $E_v^{A,B}$ denote the extreme energies of the conduction and valence bands in the material $A$ and $B$, respectively.  

The total action in the heterostructure is given\cite{lap} by  
$ {\cal S}= \sum_{A,B} \int {\cal L}_{el-EM}{\bm d}^3{\bm r} + {\cal S}_{\Pi} $.  Following the approach of Ref. \onlinecite{lap} one can  show that the contribution  ${\cal S}_{\Pi}$ of the boundary region in such a model depends only on the values of $\Psi_c$ at $z=-a$ and $z=b$. A variation of the action $\delta S=0$ with respect to  $ \Psi_c^*$ (with electromagnetic potentials $V$ and ${\bm A}$ assumed to be the constant functions of the coordinates) in a standard fashion  leads to the bulk  equation for the electron wave function $\Psi_c$, Eq.~(\ref{psic2}),  together with appropriate boundary conditions at the heterointerface. The  boundary condition parameters generally depend on the properties of the boundary region $\Pi$. For the ``ideal'' interface $|a+b|\rightarrow 0$ (see Ref. \onlinecite{lap}), they can be written as continuity conditions at $z=0$ of the conduction band spinor function $\Psi_c={\rm const}$ and of the normal projection $v_\tau= ({\bm \tau} {\bm v})={\rm const}$ (here ${\bm \tau}$ is the unit vector  normal to the interface) of the effective velocity vector ${\bm v}$:
\begin{eqnarray}
{\bm v} =  \frac{1}{m_c(\epsilon )} {\bm \pi} \Psi_c + i\, \frac{(g_c(\epsilon)-g_e)}{2m} [{\bm \sigma} \times {\bm \pi}]\Psi_c  \, .
\end{eqnarray}

The variation of the action $\delta {\cal S}=0 $ with respect to the electromagnetic potentials 
$V$ and  $A_i$, $(i=1,2,3)$ (with the wave function $\Psi_c$ assumed to be the constant functions of coordinates) leads to stationary Maxwell equations for ${\bm D}=\varepsilon {\bm E} + 4\pi {\bm \Pi}$ and ${\bm H} = {\bm B}/\mu - 4\pi {\bm M}$:
\begin{eqnarray}
{\bm \nabla}\cdot {\bm D} = 4\pi \rho_c  \, , \quad 
{\bm \nabla}\times {\bm H} = \frac{4 \pi}{c}{\bm J}_c\, 
\label{maxs2}
\end{eqnarray}
together with appropriate boundary conditions at the interface: 
\begin{eqnarray}
({\bm D}{\bm \tau}) = {\rm const} \, , \qquad ({\bm H} \times {\bm \tau}) = {\rm const} \, .
\end{eqnarray}
Keeping only the first order terms in $|\epsilon|/E_g$ and $|\mu_BB|/E_g$, we obtain expressions for the source terms $\rho_c$ and ${\bm J}_c$ and for  polarization vectors ${\bm \Pi}$ and ${\bm M}$ exactly in the same way as in Eqs. (\ref{s1}-\ref{s4}) of Sec \ref{kane}. Thus, the two approaches, the approximation of the eight-band model and the least action principle for the nonparabolic electrons, produce exactly the same results for the bulk semiconductor. 
In addition, the second approach gives boundary conditions at the interface for the envelope functions and for the electromagnetic field.  

\begin{figure}
\begin{center}
\vskip 1cm
\includegraphics[width=5.5cm,height=8cm]{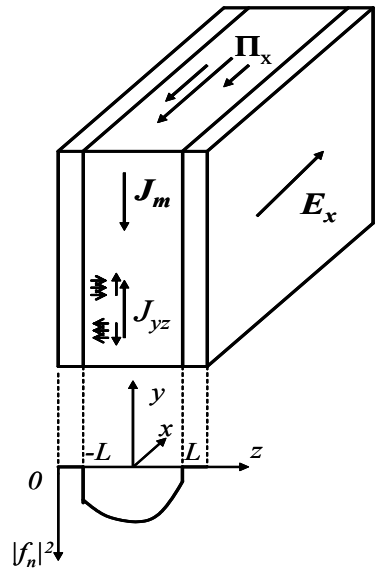}
\end{center}
\caption{\label{Fig2}  Sketch of  a square asymmetric quantum well with infinite potential barriers at $z= \pm L$. The asymmetric distribution of the electron density $|f_n(z)|^2$ and  of the electric polarization $\Pi_x(z)$ are shown schematically. The  spin Hall current ${J}_{yz}$ and the induced magnetic charge Hall current ${\bm J}_{m}$ flow in the $y$ direction when the electric field $E_x$ is applied.}
\end{figure}

\begin{widetext}

\section{Intrinsic electric polarization of the  steady state electron gas in an asymmetric quantum well}
\label{sample}

In this section we consider the spin-orbit contribution to the electric polarization which is  created by the in-plane motion of electrons in an asymmetric square quantum well.  In such  a structure,  the Rashba-type spin-orbit splitting of the electron energy levels can appear in absence of external electromagnetic field 
(${\bm E} = 0$ and ${\bm A}=0$) due to the asymmetry of the interfaces at $z=\pm L$.\cite{silva,lap,vasko} When the interfaces  are modeled by  infinite potential barriers (both in the conduction and in the valence bands), this asymmetry is  reflected by the asymmetric boundary conditions:\cite{lap}
\begin{eqnarray}
\Psi_c(\pm L) = \pm \, a^\pm\, \left. \left( \frac{m}{m_c({\cal E})}\frac{\partial \Psi_c}{\partial z} +
\frac{g_e-g_c({\cal E})}{2} [{\bm \sigma} \times {\bm k} ]_z \Psi_c \right) \right|_{z=\pm L} \, ,
\label{ibc}
\end{eqnarray}
where ${\bm k}=(k_x,k_y)$ is the wave vector of the in--plane motion, $a^\pm = a_0/t^\pm$, $a_0 = \sqrt{\hbar^2/2E_pm}$; $t^{+}$ and $t^{-}$ are real numbers. Here we consider  only the case of $t^\pm < 0$ which corresponds to the electron energy levels 
 $E_n>0$ at ${\bm k}=0$ ($n$ is the number of the electron subband). The electron wave function can be written as 
\begin{equation}
\Psi({\bm r})= \sqrt{\frac{m_c}{m_c({\cal E})}}f_n(z)\phi({\bm \rho })\, ,
\end{equation}
where 
\begin{eqnarray}
f_n(z)=C^{+}\exp(ik_nz)+C^{-}\exp(-ik_nz) \, \quad {\rm for} |z| \le L \, , 
\label{fn} \\
f_n(z)=0 \, \quad {\rm for} |z| > L \, \nonumber
\end{eqnarray}
describes the electron quantization, and $k_n=\sqrt{2m_c(E_n)E_n/\hbar^2}$. Constants  $C^\pm$ are determined by the boundary conditions given by Eq. (\ref{ibc}) together with the normalization condition $\int_{-L}^{L}|f_n|^2dz = 1$. The asymmetry of the boundary conditions (\ref{ibc}) results in the asymmetry of the electron density distribution $|f_n(z)|^2$ inside the well  as shown schematically in Fig. \ref{Fig2}. 

  The function $\phi({\bm \rho } ) =\phi (x,y)$ describes the in--plane electron motion and satisfies  the equation $\hat H_R({\cal E})\phi({\bm \rho}  )={\cal E}\phi({\bm \rho}  )$ with effective nonparabolic  Rashba Hamiltonian: 
\begin{equation}
\hat H_R({\cal E})=  \frac{\hbar^2{(k_n^2+\bm k}^2)}{2m_c({\cal E})} + \alpha_{SIA}({\cal E})[{\bm \sigma} \times {\bm k}]_z \, 
\end{equation}
with the effective coupling constant approximated as\cite{lap}
\begin{eqnarray}
\alpha_{\rm SIA}({\cal E})=\frac{\hbar^2}{4m}\frac{m_c}{m_c(E)}(g_e-g_c({\cal E})) \left( |f_n(-L)|^2-|f_n(+L)|^2 \right) \, .
\end{eqnarray}
The second term in the Hamiltonian $\hat H_R(E)$ describes  the effective spin--orbit interaction  caused by the asymmetry of the interfaces.   In presence of the external electric field ${\bm E}=(0,0,E_z)$ the effective coupling constant $\alpha_{\rm SIA}$ should be replaced by $\alpha_{\rm SIA} + \alpha_R$, where $\alpha_R = (e\hbar^2/4m) \alpha_{so} E_z$  is  the  Rashba constant. Note that in this case  functions $f_n(z)$ and the energy levels $E_n$ should be calculated taking into account the spatial dependence of the scalar potential 
$V(z)=-ezE_z$. As a result, the asymmetry of $|f_n(z)|^2$ is caused by  the asymmetry of the boundary conditions as well as by the effect of $E_z$. Although hereafter we will assume $E_z=0$,  all results  can be readily generalized for $E_z \ne 0$. 

The eigenfunctions $\psi_{\lambda, {\bm k}}({\bm \rho})$ of the Hamiltonian $\hat H_R(E)$ are given by
\begin{equation}
\phi_{{\bm k} \lambda}({\bm \rho})= \frac{e^{i{\bm k}{\bm \rho}}}{\sqrt{2S}}
\left( \begin{array}{c} 1 \\ -i \lambda k_{+}/k \end{array} \right)  \, .
\end{equation}
where $k =|{\bm k}| = \sqrt{k_x^2+k_y^2}$,  $k_{+}=k_x + i k_y$,  $S$ is the cross--section of the quantum well, and $\lambda = \pm 1$ correspond to two spin--split spectral branches. Their energies can be found from the equation 
\begin{equation}
\epsilon_{n,\lambda, k} = E_n\frac{m_c(E_n)}{m_c(\epsilon_{n,\lambda, k})} + \frac{\hbar^2 k^2}{2m_c(\epsilon_{n,\lambda, k})} + \lambda \alpha_{\rm SIA}(\epsilon_{n,\lambda, k}) k \, .
\end{equation} 

The spin-orbit interaction shifts the energy minimum of the $n$-th subband from $E_n$ to $(E_n-E_{n0})$ [see Fig. \ref{Fig3}(a)], where 
the energy shift $E_{n0}$ can be  found from the nonlinear equation
\begin{eqnarray}
E_{n0}=\frac{\alpha_{\rm SIA}^2(E_n-E_{n0})m_c(E_n-E_{n0})}{2\hbar^2} \, .
\end{eqnarray} 
In what follows we neglect the nonparabolicity of the effective mass and of  the coupling constant inside the $n$-th subband and assume $m_c(\epsilon_{n,\lambda, k}) \approx m_c(E_n)=m_n$, 
$\alpha_{\rm SIA}(\epsilon_{n,\lambda, k}) \approx \alpha_{\rm SIA}(E_n) =\alpha_n$.

 Our aim  is to calculate  the electric polarization vector:
\begin{equation}
\quad {\bm \Pi }= -\frac{ e \hbar^{2} a_p}{4m} {\bm \nabla} |\Psi_c|^2 -\frac{e\hbar^2\alpha_{so}}{8m}
 \left[\Psi_c^* {\bm \sigma }\times {\bm k }\Psi_c  -({\bm k }\Psi_c)^*\times {\bm \sigma}\Psi_c  \right] \, .
\end{equation}
In  equilibrium, we have ${\bm \Pi} = (0,0,\Pi_z)$, and for the electron state characterized by the quantum numbers $n,\lambda,{\bm k}$ we obtain:
\begin{eqnarray}
\Pi_z^{n,\lambda,{\bm k}}(z) = -\frac{ e \hbar^{2} a_p}{4m}\frac{m_c}{m_n S}\frac{\partial|f_n(z)|^2}{\partial z}
-\frac{e\hbar^2\alpha_{so}}{4m}\frac{m_c}{m_n S}|f_n(z)|^2\lambda k  \, .
  \end{eqnarray}
The polarization is inhomogeneous in the $z$ direction, and it consists of two contributions. The first contribution is due to the nonparabolicity of the electron effective mass and to the asymmetry of the electron charge density distribution in the well. It does not depend on the in--plane vector, and it is the same for both spin states of the electron. By contrast, the sign of the second contribution is opposite for branches   $\lambda = 1$ and $\lambda = -1$. It is related to the nonparabolicity of the electron effective $g$ factor. Regardless of this difference, both terms appear due the interaction of the conduction band electrons with the valence band states. 

To obtain the full polarization created in the well, one has to integrate over the equilibrium Fermi distribution corresponding  to the Fermi energy $\epsilon_F$:
\begin{eqnarray}
\Pi_z^{eq}(z)= \sum_{n,\lambda} \frac{S}{(2\pi)^2}\int  \Pi_z^{n,\lambda,{\bm k}}(z){\bm d}^2{\bm k}=\sum_n \Pi_z^n(z) \, .
\end{eqnarray}
At zero temperature $T=0$, the integration for each occupied $n$-th subband should be performed over $0<k \leq  K_{\pm}$, where $K_{\pm}$ are Fermi momenta for both spectral branches for given $n$ [see Fig. \ref{Fig3} (a)]. $K_{\pm}$ are  determined by
\begin{eqnarray}
\epsilon_F = E_n+ \frac{\hbar^2 K_{\pm }^2}{2m_n} \mp  \alpha_nK_{\pm} \, .   
\label{ef}
\end{eqnarray}
 If the Fermi level crosses only the lowest spectral branch $\lambda = -1$ of the $n$-th subband, the integration should be carried out over $K_{-} < k < K_{+}$, where 
  \begin{equation}
K_{\pm}=\frac{\alpha_nm_n}{\hbar^2} \pm \sqrt{\frac{2m_n}{\hbar^2}\left(\epsilon_F - E_n + E_{n0}\right)} \, .
\end{equation}
 
For the Fermi energies $E_1<\epsilon_F \leq E_2-E_{20}$ only the first electron subband is filled, and  both spectral branches are crossed by the Fermi level [see Fig. \ref{Fig3}(a)]. Then, the integration over $k$ and the sum over $\lambda =\pm 1$ give us the contribution from the $n=1$ subband as 
\begin{eqnarray}
\Pi_z^n(z) = -\frac{ e a_p }{4 \pi } \frac{m_c}{m} \left(\epsilon_F-E_n + 2E_{n0} \right) \frac{\partial|f_n(z)|^2}{\partial z}
+ \frac{e \alpha_n \alpha_{so}}{2 \pi \hbar^2} \frac{m_nm_c}{m} 
\left( \epsilon_F-E_n + \frac{4}{3}E_{n0} \right) |f_n(z)|^2\, .
\end{eqnarray}  
For the Fermi energies $E_1-E_{10} <\epsilon_F < E_1$ only the  lowest spectral branch of the first electron subband is filled and crossed by the Fermi level [see Fig. \ref{Fig3}(a)]. Then, integration over $k$ gives us the contribution from the $n=1$ subband as 
\begin{eqnarray}
\Pi_z^n(z) = -\frac{ e a_p\alpha_n}{4 \pi } \frac{m_c}{m}\sqrt{\frac{2m_n}{\hbar^2}\left(\epsilon_F - E_n + E_{n0}\right)} \frac{\partial|f_n(z)|^2}{\partial z}
 \\
+ \frac{e \alpha_{so}}{2 \pi } \frac{m_c}{m} \sqrt{\frac{2m_n}{\hbar^2}\left(\epsilon_F - E_n + E_{n0}\right)} 
\left( \epsilon_F-E_n+ \frac{5E_{n0}}{2} \right) |f_n(z)|^2\, . \nonumber
\end{eqnarray}  
When the Fermi energy is increased, more subbands give a contribution into polarization, and the final equilibrium polarization can be found as $\Pi_z^{eq}(z)=\sum_n \Pi_z^n(z)$.

Let us now consider the effect of the dc electric field  $E_x$ in the $x$ direction. We deal with the perturbations
 $\hat H^{(1)}=-eE_x x$ and $\hat H^{(2)}=-(e\hbar^2/4m)\alpha_{so}E_x \hat \sigma_z \hat k_y$. The second  perturbation $\hat H^{(2)}$ is related to the dependence  of the electron effective $g$--factor on $E_x$, and it describes an additional spin--orbit coupling. The perturbation related to the dependence  of the electron effective mass on $E_x$ can be neglected as far as we assume $E_x$ to be small and the size of the sample in $x$ direction to be large.  

 The  first order correction to the in--plane wave function caused   by $\hat H^{(1)}$ is 
\begin{equation}
\phi_{{\bm k} \lambda} ^{(1)} ({\bm \rho})=
\frac{\lambda e E_x k_y}{4\alpha_n k^3} \phi_{{\bm k} -\lambda}({\bm \rho}) \, .
\end{equation}
The correction to the wave function caused by $\hat H^{(2)}$ is smaller by a factor of $E_n/E_g$, and it can be neglected here.  However, we shall later consider  the first order correction to the spin--orbit energy  which corresponds to $\hat H^{(2)}$.
 
We calculate the intrinsic spin--orbit contribution to the electric polarization $\Pi_x$ which is linear in the electric field $E_x$ as 
\begin{eqnarray}
{\bm \Pi}_x(z) =   \frac{e\hbar^2\alpha_{so}}{4m}\sum_{n,\lambda} \frac{m_c}{m_n}
 |f_n(z)|^2  \frac{S}{(2\pi)^2} \int {\bm d}^2{\bm k} 
\left[ \phi^{*}_{{\bm k} \lambda} {\hat \sigma_z }{k_y } \phi_{{\bm k} \lambda} ^{(1)}  
-({k_y } \phi_{{\bm k} \lambda} )^*{\hat \sigma_z} \phi_{{\bm k} \lambda} ^{(1)}  \right]\, . 
\end{eqnarray}
The result is  ${\bm \Pi}_x(z)=\sum_n {\bm \Pi}_x^n(z)$, where 
\begin{eqnarray}
{\bm \Pi}_x^n(z) &=&  - \frac{e^2 m_c\alpha_{so}E_x}{16 m\pi } |f_n(z)|^2 \, 
\label{pix1} \\
&{\rm for}& \,\,\,\, E_n<\epsilon_F <E_{n+1}-E_{(n+1)0} \, , \nonumber
\end{eqnarray}
and 
\begin{eqnarray}
{\bm \Pi}_x^n(z) &=&  - \frac{e^2m_c\alpha_{so}E_x}{16 m\pi } \frac{\hbar^2}{m_n\alpha_n}\,
\sqrt{\frac{2 m_n}{ \hbar^2}\left(\epsilon_F - E_n + E_{n0}\,\right)} 
 |f_n(z)|^2 \, 
\label{pix2} \\
&{\rm for}&\,\,\,\, E_n-E_{n0} <E < E_n \, . \nonumber
\end{eqnarray}

Averaging over $z$, $\left< \Pi_x \right>_z = \int_{-L}^{L}\Pi_x(z)dz/2L$, and introducing an  electric susceptibility constant $\kappa_x$, we obtain:
\begin{eqnarray}
\kappa_x(\epsilon_F)  = \frac{\left< \Pi_x \right>_z}{E_x} = \sum_n \kappa_x^n(\epsilon_F) \, . 
\end{eqnarray} 
The contribution of the $n$-th subband is given by 
\begin{eqnarray}
{\kappa}_x^n(\epsilon_F) =\kappa_0=  -\, \frac{e^2 m_c\alpha_{so}}{32 m\pi L }  \, 
\label{kap0} \\
{\rm for} \,\,\,\, E_n<\epsilon_F <E_{n+1}-E_{(n+1)0} \, ,  \nonumber
\end{eqnarray} 
and 
\begin{eqnarray}
{\kappa}_x^n(\epsilon_F) =  -\, \frac{e^2m_c\alpha_{so}}{32 m\pi L  } \frac{\hbar^2}{m_n\alpha_n}\sqrt{\frac{2m_n}{\hbar^2}\left(\epsilon_F - E_n + E_{n0}\right)} 
\, \\
{\rm for} \,\,\,\, E_n-E_{n0} < \epsilon_F < E_n \, . \nonumber
\end{eqnarray}
 
The physical meaning of the finite intrinsic electric  polarization $\Pi_x$ can be understood if one considers the spin--orbit interaction of the moving electron in the external field $E_x$. This interaction is described by the perturbation $\hat H^{(2)}$ and the respective spin--orbit energy can be calculated as $E_{so}=\int {\bm d}^3{\bm r} \sum_{n{\bm k}\lambda}\left< \Psi^*|\hat H^{(2)}|\Psi \right>$. It is easy to see that this energy is given by $E_{so} = - \int{\bm d}^3{\bm r} \Pi_x(z)E_x = -\kappa_x\,E_x^2\,\Omega$, where $\Omega=2LS$ is the sample volume.

The dependence of the intrinsic electric susceptibility $\kappa_x(\epsilon_F)$ on the Fermi energy is shown in Fig. \ref{Fig3}(b). Remarkably, the electric susceptibility keeps the constant values $\kappa_x(\epsilon_F)=n\kappa_0$ when both subbands $\lambda=1$ and $\lambda = -1$ of the $n$-th quantum size band $E_n$ are crossed by the Fermi level $\epsilon_F$ and the next $E_{n+1}$ band is empty. The  value  $\kappa_0$ of Eq. (\ref{kap0}) is independent of  the subband number $n$ and of the Fermi energy $\epsilon_F$, and it is closely  related  to the universal spin Hall conductivity value $\sigma_{yx}={\cal J}_{yz}/E_x=|e|/8\pi$ obtained in Refs. \onlinecite{rashba,universal,sohall}. Here ${\cal J}_{yz}$ denotes the intrinsic spin Hall current corresponding to the flux in the positive (negative) $y$ direction of the electrons with  spin parallel (antiparallel) to $z$   when the dc electric field $E_x$ in  the $x$ direction is applied (see Fig. \ref{Fig2}). 

The above derivation of the intrinsic contribution to the electric polarization ${\Pi}_x(z)$ in the spin Hall regime allows us to predict a  new  effect of  the induced magnetic charge current ${\bm J}_m$ in the $y$ direction  when the dc electric field $E_x$ in  $x$ direction is applied (see Fig. \ref{Fig2}).  Indeed, for the steady state electron gas we have $d{\bm M}/dt =0$, where ${\bm M}$ is the magnetic polarization vector, and according to the Eq. (\ref{mpi}) 
 the induced magnetic charge current ${\bm J}_m$ can be defined as 
\begin{equation}
{\bm J}_m = - c {\bm \nabla} \times {\bm \Pi} \, .
\label{jm0}
\end{equation}
Here we  assume  that the sample is infinite in $x$ and $y$ directions, and  we do not consider edge effects. As  electric polarizations $\Pi_x(z)$ of (\ref{pix1},\ref{pix2}) are inhomogeneous in the  $z$ direction, the $y$ component of the intrinsic induced magnetic charge current is given by  $J_{my}(z) = -c \partial \Pi_x(z)/\partial z \propto \partial|f(z)|^2/\partial z$. 
Averaging over $z$, we obtain the mean  intrinsic induced magnetic charge current:
\begin{eqnarray}
 \left<J_{my}\right>_z = \int_{-L}^{L}J_{my}(z)dz/2L =  \sigma^m_{yx} E_x \, , \\
\sigma^m_{yx} (\epsilon_F)  = \frac{ \left< J_{my} \right>_z}{E_x} = \sum_n \sigma^{mn}_{yx}(\epsilon_F) \, .
\end{eqnarray} 
Here $\sigma^m_{yx}(z)$ is the average magnetic Hall conductivity summarized over all electron states. The contribution of the $n$-th subband $\sigma^{mn}_{yx}$  is given by 
\begin{eqnarray}
\sigma^{mn}_{yx}(\epsilon_F)&=& -\, \frac{e^2 m_c\alpha_{so}}{32 m\pi L} \left( |f_n(-L)|^2 - 
|f_n(L)|^2  \right) \, \\
&{\rm for}& \,\,\, \, E_n<\epsilon_F <E_{n+1}-E_{(n+1)0} \, , \nonumber
\end{eqnarray} 
and 
\begin{eqnarray}
\sigma^{mn}_{yx}(\epsilon_F)& = &  -\, \frac{e^2m_c\alpha_{so}}{32 m\pi L} \frac{\hbar^2}{m_n\alpha_n}\sqrt{\frac{2m_n}{\hbar^2}\left(\epsilon_F - E_n + E_{n0}\right)} \left( |f_n(-L)|^2 - 
|f_n(L)|^2  \right) \, \\
&{\rm for}& \,\,  \,\,
E_n-E_{n0} < \epsilon_F < E_n \, .\nonumber
\end{eqnarray}  
The dependence of the intrinsic magnetic Hall conductivity $\sigma^m_{yx} (\epsilon_F)$ on the Fermi energy is very similar to the dependence of the intrinsic electric susceptibility $\kappa_x(\epsilon_F)$ on the Fermi energy which is shown at Fig. \ref{Fig3}(b). 

\end{widetext}

\begin{figure}
\begin{center}
\vskip 1cm
\includegraphics[width=7.5cm,height=6cm]{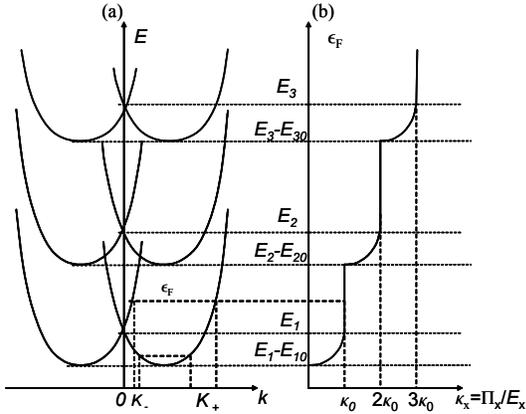}
\end{center}
\caption{\label{Fig3}  Sketch of the electron energy structure in an asymmetric quantum well (a), and the dependence of the electric susceptibility  $\kappa_x=\Pi_x/E_x$ on the Fermi energy $\epsilon_F$ in the spin Hall regime (b).}
\end{figure}

\section{Conclusion}
\label{discuss}

 In conclusion, we have considered the eight--band Kane model for the  
conduction band electrons moving in the external electromagnetic field  
and showed how the Gordon--like decomposition can be adapted to this  
setting. This approach allowed us to derive the source terms for the  
Maxwell equations and the electric and magnetic polarization vectors  
related to a moving electron. We have also derived the effective ${\bm  
k}{\bm p}$ Lagrangian for a nonparabolic conduction band and in the  
presence of the external electromagnetic field. In this way, we  
obtained boundary conditions for the envelope function and  
electromagnetic  fields at the interfaces. These results give a solid  
basis for the analysis of the spin Hall effect  and other spintronic  
effects in semiconductor heterostructures.  As an example, we have  
obtained the expression for the electric polarization induced by  the  
in--plane motion of the nonparabolic  electrons in the asymmetric  
quantum well with  infinite potential barriers and have calculated its  
dependence on the Fermi energy.  We have predicted and calculated the intrinsic 
 induced magnetic charge Hall current in the spin-Hall regime. We are going to discuss the 
relevance of our results for the fundamental questions concerning the spin Hall effect. 

First, we discuss the definition the spin Hall current. At present, three different
definitions of the spin current operator $\hat {\cal J}_{ij}$ were suggested in the literature. They are the following: (i) the conventional definition\cite{universal,rashba,dimitrova,sohall,proper,korenev} $\hat {\cal J}_{ij}=(\hbar/4)(\hat \sigma_j \hat v_i + \hat v_i \hat \sigma_j)$ (where $\hat v_i=1/\hbar(\partial \hat H/\partial k_i)$ is the velocity operator for the Hamiltonian $\hat H$); (ii) the modified definition $\hat {\cal J}_{ij}=(\hbar/2)d(\hat r_i \hat \sigma_j)/dt$ proposed in Ref. \onlinecite{def} and (iii)    
the definition $\hat {\cal J}_{ij}=-i\hbar^2/(4m_c)(\hat \sigma_j \partial_i + \partial_i \hat \sigma_j)$ obtained in Ref. \onlinecite{consistency} based on the relativistic approach. We note that the  definition (iii) ensures the relation between the anti-symmetric part of the current $\Lambda_{\alpha} = \varepsilon_{\alpha \beta \gamma}{\cal J}_{\beta \gamma}$ (here $ \varepsilon_{\alpha \beta \gamma}$ is Levi-Civita anti-symmetric tensor) and the  electric polarization $\Pi_\alpha$.\cite{consistency} It is also with this definition the universal conductance
$\sigma_{yx}=|e|/8\pi$ remains unchanged when the  $\hat H^{(2)}$ perturbation is taken into account. Finally, with this definition  the intrinsic electric polarization $\Pi_x$  is proportional to the intrinsic spin Hall current ${\cal J}_{yz}$, and the latter can be related to the spin--orbit energy $E_{so}$. This fact also plays an important role in the discussion of the cancellation (or non cancellation) of the total spin Hall current in the steady state regime. 

The steady state regime is only possible when the scattering is present in the system. In this case, the total spin Hall current ${\cal J}_{yz}$ includes contributions caused by the asymmetric scattering\cite{DP1,DP2,hirsch}, and by the generated nonequilibrium spin  density $S_y$.\cite{korenev,KKM} According to the ``cancellation theorem'' of Ref. \onlinecite{dimitrova}, the spin torque $T_y$ is proportional to ${\cal J}_{yz}$, and it vanishes in the steady state regime. However, the spin torque in Ref. \onlinecite{dimitrova} was calculated without taking into account the perturbation $\hat H^{(2)}$ which gives an additional contribution to $T_y$. This  additional contribution $1/2E_x\Pi_z^{eq}$ to the torque $T_y$ was considered in Ref. \onlinecite{consistency},
and it can be incorporated in the more general form  of the ``cancellation theorem''  presented in Ref. \onlinecite{korenev}. The direct relation between ${\cal J}_{yz}$ on the one hand, and 
$\Pi_x$ and $E_{so}$ on the other hand, implies that ${\cal J}_{yz}$ does not vanish
once the contribution to $E_{so}$ corresponding to the perturbation $\hat H^{(2)}$ is taken into account. If we assume that this energy is determined by the intrinsic properties of the system, and that it is not affected by the elastic scattering, we conclude that it is exactly the intrinsic part of the spin Hall current which remains non vanishing in the steady state regime. The extrinsic current is cancelled in the way  discussed in Ref. \onlinecite{korenev}, and it does not contribute to the total electric polarization $\Pi_x$.

We are grateful to I.A. Merkulov for the discussion of the Kane model and to V. L. Korenev for the discussion of the spin Hall current ``cancellation theorem''.  
A.V. Rodina acknowledges the support from the Swiss National Science  
Foundation.

\end{document}